\documentclass[notitlepage,amssymb,amsmath,floatfix,nofootinbib,superscriptaddress]{revtex4-1}
\usepackage{epsfig}
\usepackage{bm}
\usepackage{amssymb}
\usepackage{amsmath}
\usepackage{color}
\usepackage{xcolor}
\usepackage{slashed}
\usepackage[colorlinks,linkcolor=blue,anchorcolor=black,citecolor=blue]{hyperref}
\usepackage{ulem}
\allowdisplaybreaks[4]

\begin{document}

\title{Gluonic contributions to semi-inclusive DIS in the target fragmentation region}

\author{Kai-Bao Chen}
\email{chenkaibao19@sdjzu.edu.cn}
\affiliation{School of Science, Shandong Jianzhu University, Jinan, Shandong 250101, China}

\author{Jian-Ping Ma}
\email{majp@itp.ac.cn}
\affiliation{CAS Key Laboratory of Theoretical Physics, Institute of Theoretical Physics, P.O. Box 2735, Chinese Academy of Sciences, Beijing 100190, China}
\affiliation{School of Physical Sciences, University of Chinese Academy of Sciences, Beijing 100049, China}
\affiliation{School of Physics and Center for High-Energy Physics, Peking University, Beijing 100871, China}

\author{Xuan-Bo Tong}\email{xuan.bo.tong@jyu.fi} 
\affiliation{Department of Physics, University of Jyväskylä, P.O. Box 35, 40014 University of Jyväskylä, Finland}
\affiliation{Helsinki Institute of Physics, P.O. Box 64, 00014 University of Helsinki, Finland}

\begin{abstract}
We study one-loop contributions to semi-inclusive deep inelastic scattering in the target fragmentation region for a polarized lepton beam and nucleon target. Complete one-loop results at leading twist are derived, with a particular focus on the gluonic channel. It shows that four structure functions are generated uniquely by the gluon fracture functions starting at one-loop. Additionally, we obtain two structure functions associated with the longitudinal polarization of the virtual photon, and they are contributed by both gluon and quark channels. By combining existing twist-3 results, all eighteen structure functions for the studied process are predicted in terms of fracture functions convoluted with perturbative coefficient functions. 
\end{abstract}
\maketitle

\section{Introduction \label{sec:Introduction}}
Semi-inclusive deep inelastic scattering (SIDIS) plays an important role in hadronic physics. Experiments with 
SIDIS offer golden opportunities for probing the inner structure of the nucleon (see e.g.~\cite{Blumlein:2012bf,Boussarie:2023izj} for reviews).
The kinematic regions of SIDIS can be roughly divided into two parts~\cite{Boglione:2022gpv,Boglione:2019nwk,Gonzalez-Hernandez:2018ipj,Boglione:2016bph,Mulders:2000jt}, known as the current fragmentation region (CFR) and the target fragmentation region (TFR).
The two regions are complementary and both provide us insights into the internal structure of the nucleon, as well as the properties of strong interactions and QCD. 
In the CFR, the final state hadron moves into the forward region of the virtual photon.
One can use the conventional collinear factorization ~\cite{Collins:1989gx,Daleo:2004pn,Kniehl:2004hf,Wang:2019bvb} or transverse-momentum-dependent~(TMD) factorization~\cite{Collins:1981uk,Collins:2004nx,Ji:2004wu,Ji:2004xq,Collins:2011zzd,Rodini:2023plb,Boussarie:2023izj} to describe the process.
While in the TFR, the measured hadron predominantly travels in the forward direction of the incoming nucleon.
The concept of fracture functions~\cite{Trentadue:1993ka,Berera:1995fj,Grazzini:1997ih} was introduced for describing the factorization of the hadron production in the TFR.
The factorization with these functions has been proven to hold at leading twist  in QCD~\cite{Collins:1997sr}.

Considerable progress  has been made in studies on hadron production in the CFR.
For example, higher-order $\alpha_s$ corrections to various SIDIS structure functions in the CFR have been calculated in Refs.~\cite{Ji:2004wu,Ji:2004xq,Kang:2011mr,Sun:2013hua,Benic:2019zvg,Benic:2021gya,Benic:2024fvk,Anderle:2016kwa,Abele:2021nyo,Goyal:2023xfi,Bonino:2024qbh} within the TMD or collinear formalism, as well as in Refs.~\cite{Bergabo:2022zhe,Bergabo:2024ivx,Caucal:2024cdq} within the small-$x$ formalism.
Although not as intensively studied as those for the CFR case, there has been continuous progress in research on TFR processes.
For Drell-Yan lepton-pair production associated with one hadron in the forward- or backward regions, factorization with fracture functions have been shown to hold at one-loop level~\cite{Ceccopieri:2008fq,Ceccopieri:2010zu}.
The classification of quark TMD fracture functions for a polarized nucleon has been obtained  in~\cite{Anselmino:2011ss}, and the two-hadron production in the current and target fragmentation regions are investigated~\cite{Anselmino:2011bb,Anselmino:2011vkz}.
Studies on the factorization properties for fracture functions in different kinematic regions have also been carried out~\cite{Chai:2019ykk,Chen:2021vby,Guo:2023uis,Hatta:2022lzj,Iancu:2021rup,Iancu:2022lcw,Hauksson:2024bvv}. For instance, the small-$x$ behavior of diffractive fracture functions is studied in~\cite{Hatta:2022lzj,Iancu:2021rup,Iancu:2022lcw,Hauksson:2024bvv,Tong:2023bus,Shao:2024nor,Hatta:2024vzv}. Additionally, twist-3 contributions for SIDIS in the TFR have been calculated at large hadron transverse momentum ~\cite{Chen:2021vby} and small transverse momentum in~\cite{Chen:2023wsi}.
Moreover, energy correlators have been applied in DIS to probe the physics of TFR in recent works~\cite{Liu:2022wop,Liu:2023aqb,Li:2023gkh,Cao:2023oef,Cao:2023qat}.

On the experimental side, the observation of hadron production in the TFR has been first made in HERA experiment~\cite{ZEUS:1993vio}. Phenomenological studies based on the fracture function formalism were conducted in~\cite{Goharipour:2018yov,Khanpour:2019pzq,Maktoubian:2019ppi,Salajegheh:2022vyv,Salajegheh:2023jgi}. The CLAS Collaboration at JLab has recently reported the novel beam-spin asymmetries in the dihadron correlation which provide a first access to TMD fracture functions~\cite{CLAS:2022sqt}. A possible explanation for the asymmetries can be found in \cite{Guo:2023uis}. 
The TFR physics is expected to be studied more intensively through the potential SIDIS experiments, such as the JLab@22GeV program~\cite{Accardi:2023chb} and the planned electron-ion colliders in the USA~\cite{Boer:2011fh,Accardi:2012qut,AbdulKhalek:2021gbh,AbdulKhalek:2022hcn,Abir:2023fpo} and China~\cite{Anderle:2021wcy}.
In light of these experimental advancements, it is crucial to conduct theoretical researches in advance.

In general, there are eighteen structure functions describing polarized lepton-nucleon SIDIS with unpolarized or spin-0 hadron production~\cite{Bacchetta:2006tn}.
At twist-2 and tree-level, it has been shown that there are only four nonzero structure functions for SIDIS in the TFR~\cite{Anselmino:2011ss,Chen:2023wsi}. Only quark fracture functions contribute to these four structure functions at this order.
However, SIDIS experiments in the TFR can provide information more than that contained in the four structure functions. In order to probe more about TFR physics, one needs to make predictions about other structure functions,  and more accurately about the aforementioned four structure functions.    
With this motivation we study at leading twist the one-loop correction of SIDIS in the TFR. 
Including the one-loop correction, one expects that more structure functions would become nonzero. 
At one-loop, gluon contributions are involved. They lead to novel asymmetries which cannot be generated by quarks. This provides us a unique way to probe the gluon fracture functions and to understand the role played by gluons in the process.  Some structure functions at twist-2 have been studied beyond tree-level  in~\cite{Graudenz:1994dq,deFlorian:1995fd,Daleo:2003jf,Daleo:2003xg}, where the transverse momentum of the produced hadron is integrated over. In the current work, we derive 
one-loop contributions to all structure functions at twist-2 with the fixed transverse momentum of the final hadron.   

The rest of this paper is organized as follows.
In Sec.~\ref{sec:Kinematics}, we provide the notations and discuss the general form of the cross section in terms of the structure functions for SIDIS in the TFR. 
The tree-level structure function results are also presented in this section.
In Sec.~\ref{sec:HadronicTensor}, we present detailed one-loop calculations of the gluonic contribution to the hadronic tensor.
In Sec.~\ref{sec:SFresults}, we give the complete one-loop results for the structure functions, including  contributions from quark fracture functions.  
A short summary is given in Sec.~\ref{sec:Summary}.

\section{The notations and structure functions for SIDIS in the TFR} 
\label{Kinematics}
Through out this paper, we use the light-cone coordinate system, in which a four-vector $a^\mu$ is expressed as $a^\mu = (a^+,a^-, \vec a_\perp) = \bigl((a^0+a^3)/{\sqrt{2}}, (a^0-a^3)/{\sqrt{2}}, a^1, a^2 \bigr)$.
With the light cone vectors $n^\mu = (0,1,0,0)$ and $\bar n^\mu = (1,0,0,0)$, the transverse metric is defined as $g_\perp^{\mu\nu} = g^{\mu\nu} - \bar n^\mu n^\nu - \bar n^\nu n^\mu$, and the transverse antisymmetric tensor is given as $\varepsilon_\perp^{\mu\nu} = \varepsilon^{\mu\nu\alpha\beta} \bar n_\alpha n_\beta$ with $\varepsilon^{0123} = -\varepsilon_{0123} = 1$. We also use the notation $\tilde a_\perp^\mu \equiv \varepsilon_\perp^{\mu\nu} a_{\perp\nu}$. 

\label{sec:Kinematics}
We consider the SIDIS process with a polarized electron beam and nucleon target, i.e., $e(l,\lambda_e) + h_A (P,S) \to e(l^\prime) + h(P_h) + X$.
The momenta of the incident electron, the outgoing electron, the initial nucleon and the detected final-state hadron are denoted by $l$, $l^\prime$, $P$, and $P_h$, respectively.
At the leading order of quantum electrodynamics, there is an exchange of one virtual photon with momentum $q=l-l^\prime$ between the electron and the nucleon.
The helicity of the electron is denoted by $\lambda_e$, and $S$ is the polarization vector of the nucleon.
We consider the production of a spin-0 or unpolarized final-state hadron $h$. 
The set of Lorentz invariant variables used for SIDIS in the TFR are conventionally defined by~\cite{Graudenz:1994dq,Anselmino:2011ss,Boglione:2019nwk}
\begin{align}
 Q^2 = -q^2,~ x_B = \frac{Q^2}{2 P\cdot q},~  y=\frac{ P\cdot q}{P\cdot l},~ \xi_h = \frac{P_h \cdot q}{P \cdot q}. 
\end{align}

We work in the reference frame where the nucleon $h_A$ moves along the $z$-direction and the virtual photon moves in the $-z$-direction. 
In this frame, the momenta of the particles are given by
\begin{align} 
& P^\mu \approx ( P^+,0,0,0), \quad 
 P_h^\mu = (P_h^+, P_h^-, P_{h\perp}^1, P_{h\perp}^2), \nonumber\\
& l^\mu = \Big(\frac{1-y}{y}x_B P^+,~ \frac{Q^2}{2x_B y P^+},~ \frac{Q\sqrt{1-y}}{y},~0\Big), \quad 
 q^\mu =\Big(-x_B P^+,~ \frac{Q^2}{2x_BP^+}, 0,0\Big).
\end{align}
We use $P_{h\perp}$ to denote the length of the transverse vector $P_{h\perp}^\mu$ given by $P_{h\perp} 
\equiv \sqrt{-g_{\perp\mu\nu} P_h^\mu P_h^\nu}$. 
 For the case that the produced hadron $h$ has small transverse momentum and in the TFR, we have $P_h^+ \gg P_{h\perp} \gg P_h^-$ and $\xi_h \approx P_h^+/P^+$,
which specifies the longitudinal momentum fraction of the nucleon taken by the final-state hadron $h$.
The polarization vector of the nucleon can be decomposed by
\begin{align}
S^\mu = S_L \frac{P^+}{M} \bar n^\mu + S_\perp^\mu - S_L \frac{M}{2P^+} n^\mu,
\end{align}
where $M$ is the nucleon mass, $S_L$ denotes the longitudinal polarization of the nucleon and $S_\perp^\mu = (0,0,S_\perp^1,S_\perp^2)$ is the transverse polarization vector.
 
The incoming and outgoing electron span the lepton plane. 
We define the azimuthal angle $\phi_h$ for $ P_{h\perp}^\mu$ with respect to the lepton plane, and $\phi_S$ is that for $S_\perp^\mu$.
The azimuthal angle of the outgoing lepton around the lepton beam with respect to the spin vector is denoted by $\psi$.
In the kinematic region of SIDIS with large $Q^2$, one has $d\psi\approx d\phi_S$~\cite{Diehl:2005pc}.
With these specifications, the differential cross section is given by
\begin{align} 
\frac{ d\sigma}{dx_B dy d\xi_h d\psi d^2 P_{h\perp} } ={\frac{\alpha^2 y}{ Q^4} }  L_{\mu\nu} (l,\lambda_e,l^\prime) W^{\mu\nu}(q,P,S,P_h),
\label{eq:CrossSection}
\end{align}
where $\alpha$ is the fine structure constant.
The leptonic tensor is
\begin{align}
L^{\mu\nu}(l,\lambda_e,l^\prime)=
2 (l^\mu l^{\prime\nu} + l^\nu l^{\prime\mu} - l\cdot l^\prime g^{\mu\nu}) +  2i \lambda_e \epsilon^{\mu \nu\rho\sigma} l_\rho l^\prime_\sigma.
\label{eq:LeptonicTensor}
\end{align}
The hadronic tensor is defined by
\begin{align} 
W^{\mu\nu}(q,P,S,P_h) = {\frac{1}{4\xi_h} }\sum_X  \int \frac {d^4 x}{(2\pi)^4} e^{iq\cdot x} \langle S;h_A \vert J^\mu (x) \vert h X\rangle \langle X h \vert J^\nu (0) \vert h_A;S \rangle,
\label{eq:Wuv}
\end{align}
where $J^\mu(x) = \sum_f e_f \bar \psi(x) \gamma^\mu \psi(x)$ is the electromagnetic current with $f$ for all flavors.
Contracting the general form of the hadronic tensor from kinematic analysis with the leptonic tensor in Eq.~(\ref{eq:LeptonicTensor}), one can get the differential cross section in terms of the structure functions. 
The general form of the differential cross section can be expressed by eighteen structure functions as follows:~\cite{Bacchetta:2006tn,Chen:2023wsi}
\begin{align}
& \frac{ d\sigma}{d x_B d y d \xi_h  d\psi d^2 P_{h\perp} } =\frac{\alpha^2}{x_B y Q^2} \Bigl\{ A(y) F_{UU,T} + E(y) F_{UU,L} + B(y) F_{UU}^{\cos\phi_h} \cos\phi_h + E(y) F_{UU}^{\cos2\phi_h} \cos2\phi_h \nonumber\\
& + \lambda_e D(y) F_{LU}^{\sin\phi_h} \sin\phi_h + S_L \Bigl[ B(y) F_{UL}^{\sin\phi_h} \sin\phi_h + E(y) F_{UL}^{\sin2\phi_h} \sin2\phi_h \Bigr] + \lambda_e S_L \Bigl[ C(y) F_{LL} + D(y) F_{LL}^{\cos\phi_h} \cos\phi_h \Bigr] \nonumber\\
& + |\vec S_\perp| \Bigl[ \bigl( A(y) F_{UT,T}^{\sin(\phi_h-\phi_S)} + E(y) F_{UT,L}^{\sin(\phi_h-\phi_S)} \bigr) \sin(\phi_h-\phi_S) + E(y) F_{UT}^{\sin(\phi_h+\phi_S)} \sin(\phi_h+\phi_S) \nonumber\\
& \qquad + B(y) F_{UT}^{\sin\phi_S} \sin\phi_S + B(y) F_{UT}^{\sin(2\phi_h-\phi_S)} \sin(2\phi_h-\phi_S) + E(y) F_{UT}^{\sin(3\phi_h-\phi_S)} \sin(3\phi_h-\phi_S) \Bigr] \nonumber\\
& + \lambda_e |\vec S_\perp| \Bigl[ D(y) F_{LT}^{\cos\phi_S} \cos\phi_S + C(y) F_{LT}^{\cos(\phi_h-\phi_S)} \cos(\phi_h-\phi_S) + D(y) F_{LT}^{\cos(2\phi_h-\phi_S)} \cos(2\phi_h-\phi_S) \Bigr] \Bigr\}.
\label{eq:SFs-SIDIS}
\end{align}
Here the functions of $y$ defined for convenience are given by
\begin{align}
& A(y) = y^2-2y+2, ~~~ B(y) = 2(2-y)\sqrt{1-y}, \quad
 C(y) = y(2-y), \nonumber\\  &D(y) = 2y\sqrt{1-y}, \quad
 E(y) = 2(1-y).
\end{align}
It should be noted that if one works in $D$-dimension with $D = 4-2\epsilon$, there would be terms proportional to $\epsilon$ in the definition of the $y$-functions, e.g.~$A(y) = y^2-2y+2-\epsilon y^2$.
All the structure functions in Eq.~(\ref{eq:SFs-SIDIS}) are scalar functions depending on $x_B$, $\xi_h$, $Q^2$ and $ P_{h\perp}^2$.
The first and second subscripts of the structure functions denote the polarization of the electron and the nucleon, respectively.
The third subscript, if any, specifies the polarization of the virtual photon.

At tree level, the structure functions only receive contributions from quark fracture functions.
At the leading twist, it has been shown that there exist four nonzero structure functions.
They are given by~\cite{Anselmino:2011bb,Chen:2023wsi}
\begin{align}
& F_{UU,T} = x_B u_{1}(x_B,\xi_h, P_{h\perp}), \quad 
 F_{LL} = x_B l_{1L}(x_B,\xi_h, P_{h\perp}), \nonumber\\
& F_{UT,T}^{\sin(\phi_h-\phi_S)} = \frac{ P_{h\perp}}{M} x_B u_{1T}^{h}(x_B,\xi_h, P_{h\perp}), \quad  F_{LT}^{\cos(\phi_h-\phi_S)} = \frac{P_{h\perp}}{M} x_B l_{1T}^{h}(x_B,\xi_h, P_{h\perp}), \label{eq:FLO}
\end{align}
where $u_{1}$, $l_{1L}$, $u_{1T}^{h}$, and $l_{1T}^{h}$ are twist-2 quark fracture functions.
They are defined through the quark fracture matrix as~\cite{Chen:2023wsi}
\begin{align}
{\cal M}_{ij}(x) &=\int \frac{d\eta^-}{2\xi_h(2\pi)^4} e^{-ixP^+\eta^-} \sum_X \langle h_A(P)|\bar \psi_j(\eta^-) {\cal L}_n^{\dagger}(\eta^-) |hX \rangle  \langle Xh| {\cal L}_n(0) \psi_i(0) |h_A(P) \rangle \nonumber\\
&=\frac{(\gamma^-)_{ij} }{2N_c} \Bigl( u_{1} - \frac{P_{h\perp} \cdot \tilde S_\perp}{M} u_{1T}^{h} \Bigr)
- \frac{(\gamma^- \gamma_5)_{ij} }{2N_c} \Bigl( S_L l_{1L}  - \frac{P_{h\perp} \cdot S_\perp}{M} l_{1T}^{h} \Bigr) + \cdots,
\end{align}
where $\cdots$ denote the contributions beyond twist-2 or the chirality-odd parts. The chirality-odd parts 
can not contribute because of the helicity conservation in perturbative QCD. At one-loop level, the gluon fracture functions are involved. These gluon contributions not only produce corrections to the leading-order results, but also give rise to new types of structure functions.
We shall focus on the gluon contributions in the next section.

\section{The hadronic tensor at one-loop}
\label{sec:HadronicTensor}
\subsection{The gluonic contribution to the hadronic tensor}
\begin{figure}[!htb]
\centering
\includegraphics[width=0.8\textwidth]{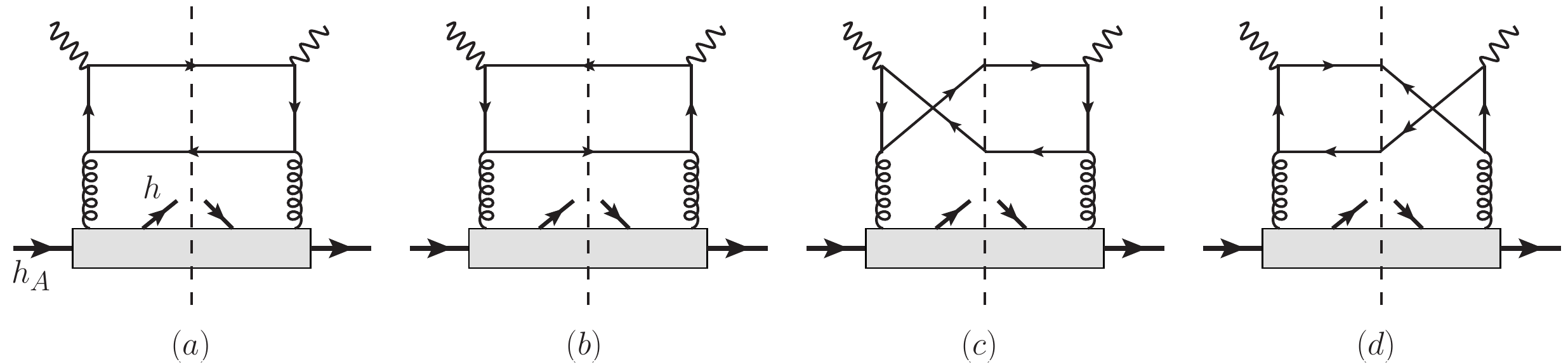}
\caption{Gluonic contributions to the hadronic tensor.}
\label{fig:gluonchannel}
\end{figure}

Now we calculate in perturbation theory for the hadronic tensor at one-loop.
We focus on the gluon channels which are shown by diagrams in Figs.~\ref{fig:gluonchannel}(a)-(d).
The gray boxes represent the gluonic correlators with a hadron $h$ identified in the final state.
These contributions can be summarized as the following form:
\begin{align}
W^{\mu \nu}(q,P,S,P_h)={\alpha_s T_F } \sum_f e_f^2   
\int \frac{d x}{x} \int d\Phi_{k_1 k_2} H^{\mu\nu \alpha\beta }(k_g, k_1, k_2)
 \mathcal{M}_{G,\alpha \beta}(x,\xi_h,P_{h\perp}),
 \label{eq:Wcon}
\end{align}
where $T_F = 1/2$. We have applied the collinear approximation where the gluon momentum obeys $k_g^\mu\approx (xP^+,0,0,0)$.
The momenta of the quark and anti-quark in the final states are denoted by $k_1$ and $k_2$, respectively.
The symbol $\int d\Phi_{k_1 k_2}$ stands for the two-body phase space integral over $k_1$ and $k_2$, i.e.,
\begin{align}
\int d\Phi_{k_1 k_2} \equiv \int \frac{d^D k_1}{(2\pi)^D} \int \frac{d^D k_2}{(2\pi)^D}(2\pi)\delta(k_1^2) (2\pi)\delta(k_2^2)(2\pi)^4 
\delta^{(4)}(q+ k_g-k_1-k_2)
\theta(q^0+ k_g^0). 
\label{eq:PhaseInt}
\end{align}
We use the dimensional regularization with $D = 4-2\epsilon$.
The hard scattering functions $H^{\mu\nu \alpha\beta }(k_g, k_1, k_2)$ in Eq.~(\ref{eq:Wcon}) receive contributions from Figs.~\ref{fig:gluonchannel}(a)-(d).
We have $H^{\mu\nu \alpha\beta} = H_{(a)+(b)}^{\mu\nu\alpha\beta} + H_{(c)+(d)}^{\mu\nu\alpha\beta}$, and
\begin{align}
H_{(a)+(b)}^{\mu\nu \alpha\beta }(k_g, k_1,k_2)= &{\rm Tr}
 \Big[ \slashed k_1 \gamma^\nu \frac{(\slashed k_g- \slashed k_2) }{( k_g- k_2)^2}\gamma^\beta \slashed k_2 \gamma^\alpha \frac{(\slashed k_g- \slashed k_2) }{( k_g- k_2)^2}\gamma^\mu \Big]+(k_1 \leftrightarrow k_2), \nonumber\\ 
 H_{(c)+(d)}^{\mu\nu \alpha\beta }(k_g, k_1,k_2)=& {\rm Tr}
 \Big[ \slashed k_1 \gamma^\nu \frac{(\slashed k_g- \slashed k_2) }{( k_g- k_2)^2}\gamma^\beta \slashed k_2 \gamma^\mu \frac{( \slashed k_1-\slashed k_g) }{(  k_1-k_g)^2}\gamma^\alpha \Big]+(k_1 \leftrightarrow k_2).
 \label{eq:H}
\end{align}
${\cal M}_G^{\alpha \beta}(x, \xi_h, P_{h\perp})$ is the gluonic correlation function defined by:
\begin{align}
{\cal M}_G^{\alpha \beta}(x, \xi_h, P_{h\perp})= & \frac{1}{2\xi_h(2\pi)^3} \frac{1}{x P^{+}} \int \frac{d \lambda}{2 \pi} e^{-i \lambda x P^{+}}\sum_X\big\langle h_A(P)\big|(G^{+\alpha}(\lambda n) {\cal L}_n^{\dagger}(\lambda n))^a\big| X h(P_{h})\big
\rangle \notag \\ 
& \times \big\langle h(P_{h}) X\big|\big({\cal L}_n(0) G^{+\beta}(0)\big)^a\big| h_A(P)\big\rangle,
\label{eq:gluonMG}
\end{align}
where $\alpha$ and $\beta$ are both transverse indices. $G^{a, \mu\nu}$ is the gluon strength tensor, where the color index $a$ ranges from 1 to 8.
${\cal L}_n (x)$ is the light-cone gauge link defined by
\begin{align}
{\cal L}_n (x) = {\cal P} \exp \biggr \{ - i g_s \int_0^\infty d\eta ~A^{+}(\eta n +x)  \biggr \},
\end{align}
where $A^\mu = A^\mu_c t^c$ with $t_{ab}^c = -if_{abc}$ is the gluon potential in the adjoint representation.
We discuss the decomposition and parametrization of ${\cal M}_G^{\alpha \beta}$ later in Sec.~\ref{sec:parametrization}.

\subsection{The results for the hard scattering functions}
To evaluate the explicit form of the hard scattering functions and the phase space integral, it is convenient to take the photon-gluon center-of-mass frame, where the initial gluon moves 
along the $z$-direction with $k_g^+=-q^+/z$ and $z=x_B/x$.  Then the relevant momenta scalar products can be written as 
\begin{align}
& q\cdot k_g=\frac{Q^2}{2 z},~~q\cdot k_1=\frac{Q^2}{4 z}(1+\cos \theta)-\frac{Q^2}{2},~~q\cdot k_2=\frac{Q^2}{4 z}(1-\cos \theta)-\frac{Q^2}{2}, \nonumber\\ 
& k_g\cdot k_2=\frac{Q^2}{4z}(1+\cos \theta),~~k_g\cdot k_1=\frac{Q^2}{4 z}(1-\cos \theta),~~k_1\cdot k_2=Q^2\frac{\bar z}{2 z}, \label{eq:SP2}
\end{align}
where $\bar z = 1-z$, and $\theta$ is the polar angle of the final quark with the momentum $k_1$.
We also notice that the light-cone vectors $\bar n^\mu$ and $n^\mu$ can be constructed by
\begin{align}
\bar n^\mu=-\frac{z}{q^+} k_g^\mu,\quad n^\mu=-\frac{2q^+}{Q^2}(q^\mu+z k_g^\mu).
\label{eq:nnbar}
\end{align}
Then, the phase space integral in Eq.~(\ref{eq:PhaseInt}) can be further simplified and evaluated as
 \begin{align}
\int d\Phi_{k_1 k_2}
=&\frac{1}{(2\pi)^{2-2\epsilon} } \frac{1}{8(\frac{\bar z}{4z}Q^2)^{2\epsilon}}   \int^{\pi}_0 d\theta (\sin \theta)^{1-2\epsilon } \int d \Omega_{\rm T}~,
\end{align}
with the normalization of the azimuthal angle integral given by $\int d \Omega_{\rm T}=2\pi^{1-\epsilon}/\Gamma(1-\epsilon)$.

To calculate the hard scattering functions, we divide them into two categories based on the polarization states of the virtual photon. Specifically, we express them as follows:
\begin{align}
H^{\mu\nu\alpha \beta} = H_L^{\mu\nu \alpha \beta } + H_T^{\mu\nu \alpha \beta }, \label{eq:H=HL+HT}
\end{align}
where
\begin{align}
 H_L^{\mu\nu \alpha \beta }
=\bar n^\mu \bar n^\nu H^{++ \alpha\beta} + n^\mu  n^\nu H^{--\alpha\beta} + \bar n^{\{\mu} n^{\nu\}} H^{+-\alpha\beta}, \quad  H_T^{\mu\nu \alpha \beta }
=g_\perp^{\mu\mu^\prime}g_\perp^{\nu\nu^\prime} H_{\mu^\prime \nu^\prime}^{~~~~\alpha\beta}. \label{eq:HT-index}
\end{align}
Note that both $\alpha$ and $\beta$ in Eqs.~(\ref{eq:H=HL+HT}) and (\ref{eq:HT-index}) are transverse indices.
The calculation of $H_L^{\mu\nu\alpha\beta}$ and $H_T^{\mu\nu\alpha\beta}$ is straightforward.
Utilizing Eq.~(\ref{eq:nnbar}), we obtain
\begin{align}
H_L^{\mu\nu\alpha\beta} =
& 32z^2(\csc\theta)^2 \left[ \frac{\bar n^\mu \bar n^\nu}{(q^-)^2} + \frac{n^\mu n^\nu}{(q^+)^2} + \frac{2\bar n^{\{\mu} n^{\nu\}}}{Q^2} \right] k_{2\perp}^\alpha k_{2\perp}^\beta, \nonumber\\
H_T^{\mu\nu \alpha \beta}=
& \frac{(\csc\theta)^4}{Q^4} \biggl\{8 Q^2\Big[8 zk_{2 \perp}^\beta(\cos ^2\theta k_{2 \perp}^\nu g^{\mu \alpha}_\perp+2 z k_{2 \perp}^\alpha g^{\mu \nu}_\perp)
\notag \\ 
&-8 z k_{2 \perp}^\nu k_{2 \perp}^\alpha g^{\mu \beta}_\perp+Q^2 \sin ^2\theta(\cos ^2\theta g^{\mu \alpha}_\perp g^{\nu \beta}_\perp-g^{\mu \beta}_\perp g^{\nu \alpha}_\perp+g^{\mu \nu}_\perp g^{\beta \alpha}_\perp)\Big] \nonumber\\
&-64 z k_{2 \perp}^\mu\Big[Q^2 k_{2 \perp}^\beta g^{\nu \alpha}_\perp+k_{2 \perp}^\alpha(8 z k_{2 \perp}^\nu k_{2 \perp}^\beta-Q^2 \cos ^2\theta g^{\nu \beta}_\perp)\Big]\biggr\}. \label{eq:HT}
\end{align}
With these results for $H^{\mu\nu \alpha \beta}$, upon integrating over the phase space, we obtain:
\begin{align}
\int d\Phi_{k_1 k_2} & H^{\mu\nu \alpha \beta} = - g_\perp^{\alpha\beta} \frac{2z \bar z}{\pi Q^2}\bar q^\mu \bar q^\nu \nonumber\\
& + \biggl(\frac{16 \pi z }{Q^2\bar z}\biggr)^\epsilon\frac{- \pi^{\epsilon-\frac{1}{2}}}{2(2-\epsilon)(1-\epsilon) \epsilon \Gamma\left(\frac{3}{2}-\epsilon\right)}\bigg\{
\frac{1}{2}(\epsilon-2)(\epsilon-1)\Bigl[g_\perp^{\alpha \mu} g_\perp^{\beta \nu}+(2 \epsilon-1)\left(g_\perp^{\alpha \nu} g_\perp^{\beta \mu}-g_\perp^{\alpha \beta} g_\perp^{\mu \nu}\right)\Bigr] \nonumber\\
& \qquad\qquad + \bar z^2 \epsilon\left(g_\perp^{\alpha \nu} g_\perp^{\beta \mu}+g_\perp^{\alpha \mu} g_\perp^{\beta \nu}+g_\perp^{\alpha \beta} g_\perp^{\mu \nu}\right) + \bar z(\epsilon-2)\Bigl[g_\perp^{\alpha \mu} g_\perp^{\beta \nu}+(2 \epsilon-1)\left(g_\perp^{\alpha \nu} g_\perp^{\beta \mu}-z g_\perp^{\alpha \beta} g_\perp^{\mu \nu}\right)\Bigr]
\bigg\},
\label{eq:H-int}
\end{align}
where $\bar q^\mu \equiv q^\mu+ 2x_B P^+\bar n^\mu$.
The first term on the right hand side of Eq.~(\ref{eq:H-int}) originates from the contribution of $H_L^{\mu\nu \alpha \beta}$. The second term arises from $H_T^{\mu\nu \alpha \beta}$, and it has a collinear divergence represented by the pole in $\epsilon$. This divergence can be subtracted, as discussed in the next section. 
  

\subsection{Parametrization of the correlation function and the gluon fracture functions}
\label{sec:parametrization}

The gluonic correlation function ${\cal M}_G^{\alpha \beta}$ defined in Eq.~(\ref{eq:gluonMG}) obeys constraints from parity and hermiticity. Its parametrization closely resembles that of gluon TMD PDFs, see e.g., Eq.~(2.141) in the TMD handbook~\cite{Boussarie:2023izj}.
In dimensional regularization with $D=4-2\epsilon$, the parametrization takes the form:
\begin{align}
\mathcal{M}_G^{\alpha \beta}= &
- \frac{1}{2-2\epsilon}g_\perp^{\alpha \beta} u_{1g} + \frac{1}{2M^2}\Big(P_{h\perp}^\alpha P_{h\perp}^\beta + \frac{1}{2-2\epsilon}g_\perp^{\alpha \beta} P_{h\perp}^2\Big) t_{1g}^{h} 
+ S_L\bigg[i\frac{\epsilon_\perp^{\alpha \beta}}{2} l_{1gL}
+ \frac{\tilde  P_{h\perp}^{\{\alpha} P_{h\perp}^{\beta\}}}{4 M^2} t_{1gL}^{h}\bigg] \nonumber\\
& + \frac{g_\perp^{\alpha \beta} }{2-2\epsilon}
\frac{P_{h\perp}\cdot \tilde S_\perp }{M} u_{1g T}^{h} + \frac{P_{h\perp} \cdot S_\perp}{M} \bigg[ i\frac{\epsilon_\perp^{\alpha \beta}}{2} l_{1gT}^h 
- \frac{\tilde P_{h\perp}^{\{\alpha} P_{h\perp}^{\beta\}}}{4M^2} t_{1gT}^{hh} \bigg]
+\frac{\tilde P_{h\perp}^{\{\alpha} S_\perp^{\beta\}}+ \tilde S_\perp^{\{\alpha} P_{h\perp}^{\beta\}}}{8 M} t_{1gT}^h + \cdots.
\label{eq:gluonFrF}
\end{align}
Here, $\cdots$ represent power-suppressed terms beyond the twist-2 level. The shorthand notation $a^{\{\alpha} b^{\beta\}} \equiv a^\alpha b^\beta + a^\beta b^\alpha$ is used. We adopt the HVBM scheme~\cite{tHooft:1972tcz,Breitenlohner:1977hr} where the Levi-Civita tensor $\varepsilon^{\mu\nu\alpha\beta}$ is a genuinely four-dimensional object and its components vanish in all unphysical dimensions.

The $u$'s, $l$'s, and $t$'s are scalar functions of $x$, $\xi_h$, and $P_{h\perp}^2$, referred to as gluon fracture functions.
Here, We adopt naming conventions akin to those used for quark fracture functions~\cite{Chen:2023wsi}. 
The subscript ``$1g$" denotes the leading twist and gluon case, while ``$L$" or ``$T$" indicates dependence on nucleon longitudinal or transverse polarizations. The symbol "h" appears in the superscript when there is an explicit dependence on the transverse momentum of the final-state hadron $h$ in the decomposition.

Specifically, $u_{1g}$ and $u^h_{1gT}$ represent the unpolarized gluon fracture functions for the unpolarized and transversely polarized nucleons, respectively. $l_{1gL}$ and $l_{1gT}^h$ denote the circularly polarized gluon fracture functions of the longitudinally and transversely polarized nucleons, respectively. $t_{1g}^h$ can be understood as an analogy to the so-called linearly polarized gluon distribution in an unpolarized nucleon, and we notice that a novel method based on the nucleon energy-energy correlator~\cite{Liu:2022wop} for probing the linearly polarized gluon has been investigated recently~\cite{Li:2023gkh}.  We  have six gluon fracture functions associated with the nucleon polarization.

\section{Results of structure functions up to one loop}
\label{sec:SFresults}
Substituting Eqs.~(\ref{eq:H-int}) and (\ref{eq:gluonFrF}) into Eq.~(\ref{eq:Wcon}), we obtain the gluon contributions to the hadronic tensor.
By further contracting with the leptonic tensor given in Eq.(\ref{eq:LeptonicTensor}) and comparing with the cross section expressed by structure functions in Eq.~(\ref{eq:SFs-SIDIS}), we obtain the final results for the gluon contributions to the structure functions, presented as follows.

We first present the results of structure functions which are zero at tree-level.  There are four structure functions, which are generated by the gluon contributions  only at ${\cal O}(\alpha_s)$.
They are: 
\begin{align}
& F_{UU}^{\cos 2\phi_h} =  -\frac{\alpha_s T_F}{2\pi}  \frac{P_{h\perp}^2}{2M^2} x_B \sum_{q,\bar q} e_q^2
\int^1_{x_B/\bar \xi_h} \frac{d z}{z} z^2 t^h_{1g}(x_B/z,\xi_h,P_{h\perp}),
 \nonumber\\
 & F_{UL}^{\sin 2\phi_h} = \frac{\alpha_sT_F}{2\pi} \frac{P_{h\perp}^2}{2M^2} x_B\sum_{q,\bar q} e_q^2
\int^1_{x_B/\bar \xi_h} \frac{d z}{z}  z^2 t^h_{1gL}(x_B/z,\xi_h,P_{h\perp}),\nonumber
 \\ 
& F_{UT}^{\sin (3\phi_h-\phi_s)} = \frac{\alpha_sT_F}{2\pi} \frac{P_{h\perp}^3}{4M^3} x_B\sum_{q,\bar q} e_q^2
\int^1_{x_B/\bar \xi_h} \frac{d z}{z} z^2 t^{hh}_{1gT}(x_B/z,\xi_h,P_{h\perp}), \nonumber
 \\ 
& F_{UT}^{\sin (\phi_h+\phi_s)} = \frac{\alpha_sT_F}{2\pi} \frac{ P_{h\perp}}{2M} x_B\sum_{q,\bar q} e_q^2  
\int^1_{x_B/\bar \xi_h} \frac{d z}{z} z^2\bigg[t^h_{1gT}(x_B/z,\xi_h,P_{h\perp}) + \frac{ P_{h\perp}^2}{2M^2}t^{hh}_{1gT}(x_B/z,\xi_h,P_{h\perp}) \bigg], \label{eq:SF-FUTsinphi+phis}
\end{align}
where $\bar \xi_h = 1-\xi_h$.
 The limits on the integration are imposed by the kinematic constraint $x_B<x=x_B/z<1-\xi_h$. These structure functions in Eqs.(\ref{eq:SF-FUTsinphi+phis}) cannot be generated by the twist-2 quark fracture functions even at one-loop. Thus,
they are uniquely generated from gluon contributions.
 These structure functions give rise to four kinds of azimuthal asymmetries, two of which depend on the nucleon transverse polarization.
At leading twist, quark fracture functions can not contribute to these structure functions because of angular 
momentum conservation. 
We note in particular that, if one uses  tree-level approximation, all these four structure functions will only have contributions beyond twist-3~\cite{Chen:2023wsi}.

Besides the four structure functions given in the above, there exist other two structure functions, namely 
 $F_{UU,L}$ and $F_{UT,L}^{\sin(\phi_h - \phi_S)}$,  generated 
not only by the gluon contributions but also by the quark contributions. It is well known in inclusive DIS that the quark distributions also contribute to $F_{UU,L}$ at one-loop. The calculation here for the TFR SIDIS goes through the same procedure.
Therefore, the hard scattering coefficient functions here for TFR SIDIS  coincide with those for inclusive DIS up to overall kinematic factors absorbed in the definition of the fracture functions.
The relevant results can be found, e.g. in~\cite{Ellis:1996mzs}.
The same principle applies to the nucleon transverse polarization-dependent counterpart $F_{UT,L}^{\sin(\phi_h - \phi_S)}$.
We have:
\begin{align}
& F_{UU,L} = \frac{\alpha_s}{2\pi} x_B  \sum_{q,\bar q} e_q^2 \int^1_{x_B/\bar \xi_h} \frac{d z}{z}\bigg[
 4T_Fz\bar z u_{1g}(x_B/z,\xi_h,P_{h\perp}) +2 C_F z u_{1}(x_B/z,\xi_h,P_{h\perp}) \bigg], \nonumber\\ 
& F_{UT,L}^{\sin \left(\phi_h-\phi_S\right)}
= \frac{\alpha_s}{2\pi} \frac{ P_{h\perp} }{M} x_B \sum_{q,\bar q} e_q^2\int^1_{x_B/\bar \xi_h} \frac{d z}{z}\bigg[
 4T_F z \bar z u_{1gT}^h(x_B/z,\xi_h,P_{h\perp}) + 2C_F z u_{1T}^h(x_B/z,\xi_h,P_{h\perp}) \bigg]. \label{eq:FUULT}
\end{align}
We note that, at tree-level, these two structure functions only have nonzero contributions beyond twist-3~\cite{Chen:2023wsi}.
However, there are indications suggesting that the ratio of the longitudinal to transverse cross sections of SIDIS,  i.e, $ F_{UU,L}/F_{UU,T}$, could be sizeable (see the discussions in Sec. 5.2 of~\cite{Accardi:2023chb} and reference therein).
We see from Eq.~(\ref{eq:FUULT}) that, besides the kinematic power suppressed effects, the one-loop  corrections should also be taken into account properly to provide a more comprehensive explanation of this ratio especially in the TFR.

Now we turn to one-loop contributions to the four structure functions in Eq.(\ref{eq:FLO}) which are already nonzero at tree-level. 
At one-loop they receive both quark and gluon channels contributions.  
For these contributions, one needs to subtract the collinear divergences because they are already included in parton fracture functions. The subtraction is similar to that in inclusive DIS.  
After the subtraction of these divergences, we obtain finite contributions to the structure functions.
Again, these finite contributions can also be found in literature, e.g., in~\cite{Ellis:1996mzs} for the unpolarized parton case, and in~\cite{Gluck:1995yr,Vogelsang:1995vh,deFlorian:1997ie} for the polarized parton case. We note that for the polarization dependent case, the coefficients generally depend on the $\gamma_5$-scheme in $D$-dimensions. {In our calculations, as mentioned in Sec. \ref{sec:parametrization}, we utilize the HVBM scheme~\cite{tHooft:1972tcz,Breitenlohner:1977hr}}.
We include the quark contributions for completeness and present the full results of these four structure functions.
They are given by
\begin{align}
& F_{UU,T} = x_B \sum_{q,\bar q} e_q^2\int_{x_B/\bar\xi_h}^1 \frac{dz}{z} \Bigl[ {\cal H}_g(z) u_{1g}(x_B/z, \xi_h, P_{h\perp}) + {\cal H}_q(z) u_1(x_B/z, \xi_h, P_{h\perp}) \Bigr], \nonumber\\
& F_{UT,T}^{\sin(\phi_h - \phi_S)} = \frac{P_{h\perp} }{M} x_B \sum_{q,\bar q} e_q^2\int_{x_B/\bar\xi_h}^1 \frac{dz}{z} \Bigl[ {\cal H}_g(z) u_{1gT}^h(x_B/z, \xi_h, P_{h\perp}) + {\cal H}_q(z) u_{1T}^h(x_B/z, \xi_h, P_{h\perp}) \Bigr], \nonumber\\
& F_{LL} = x_B \sum_{q,\bar q} e_q^2\int_{x_B/\bar\xi_h}^1 \frac{dz}{z} \Bigl[ \Delta {\cal H}_g(z) l_{1gL}(x_B/z, \xi_h, P_{h\perp}) + \Delta{\cal H}_q(z) l_{1L}(x_B/z, \xi_h, P_{h\perp}) \Bigr], \nonumber\\
& F_{LT}^{\cos(\phi_h - \phi_S)} = \frac{P_{h\perp} }{M} x_B \sum_{q,\bar q} e_q^2 \int_{x_B/\bar\xi_h}^1 \frac{dz}{z} \Bigl[   \Delta{\cal H}_g(z) l_{1gT}^h(x_B/z, \xi_h, P_{h\perp}) +\Delta{\cal H}_q(z) l_{1T}^h(x_B/z, \xi_h, P_{h\perp})\Bigr],
\end{align}
where we have suppressed the renormalization scale $\mu$-dependence in the arguments of the hard coefficient functions and the fracture functions.
The hard coefficient functions are given by
\begin{align}
& {\cal H}_q(z) = \delta(\bar z) + \frac{\alpha_s}{2\pi} \Biggl\{ P_{qq}(z) \ln\frac{Q^2}{\mu^2} + C_F \Biggl[ 2\left( \frac{\ln\bar z}{\bar z} \right)_+ - \frac{3}{2}\left(\frac{1}{\bar z}\right)_+  - (1+z)\ln\bar z - \frac{1+z^2}{\bar z} \ln z + 3 - \left( \frac{\pi^2}{3} + \frac{9}{2} \right) \delta(\bar z) \Biggr] \Biggr\}, \nonumber\\
& \Delta {\cal H}_q(z) = \delta(\bar z) + \frac{\alpha_s}{2\pi} \Biggl\{\Delta P_{qq}(z) \ln\frac{Q^2}{\mu^2} +  C_F\Biggl[ (1+z^2)\left( \frac{\ln\bar z}{\bar z} \right)_+ - \frac{3}{2}\left(\frac{1}{\bar z}\right)_+ - \frac{1+z^2}{\bar z} \ln z + 2+z - \left( \frac{\pi^2}{3} + \frac{9}{2} \right) \delta(\bar z) \Biggr] \Biggr\}, \nonumber\\
& {\cal H}_g(z) = \frac{\alpha_s}{2\pi} \Biggl[ P_{qg}(z) \ln \frac{Q^2\bar z}{\mu^2 z}  - T_F (1-2z)^2 \Biggr], \quad 
 \Delta{\cal H}_g(z) = \frac{\alpha_s}{2\pi} \Biggl[\Delta P_{qg}(z) \left( \ln\frac{Q^2\bar z}{\mu^2 z} -1 \right) + 2T_F \bar z \Biggr],
\end{align}
with the lowest-order parton splitting functions~\cite{Altarelli:1977zs}
\begin{align}
& P_{qq}(z) = C_F \left[ \frac{1+z^2}{(1- z)_+} + \frac{3}{2}\delta(1-z) \right]~, \quad
 P_{qg}(z)=T_F\left [  z^2
 +(1-z)^2\right]~, \nonumber\\
& \Delta P_{qq}(z) = P_{qq}(z)~, \quad
 \Delta P_{qg}(z)=T_F(2z-1)~.
\end{align}

Our results at the leading twist show that there are 6 structure functions which become nonzero at one-loop. 
The four structure functions which are nonzero at tree-level, receive one-loop corrections. The remaining 8 structure functions of the total 18 structure functions are zero at leading twist or twist-2. They become  nonzero when twist-3 contributions are included~\cite{Chen:2023wsi}. With the twist-3 contributions and the 
results presented in this work, all 18 structure functions of SIDIS in the TFR are predicted 
as convolutions of perturbative coefficient functions with fracture functions.

\section{summary}
\label{sec:Summary}
We have studied one-loop contribution for SIDIS in the TFR at leading twist.  Complete information of  the studied process is encoded in eighteen structure functions. At tree-level and leading twist there are only four structure functions  which are nonzero and associated with quark fracture functions. At one-loop
other six structure functions become nonzero. Special attention is paid to the gluon channel.  It is interesting to note that four of the six structure functions receive contributions only from gluon fracture functions. 
Hence,  it is important to measure these four structure functions for understanding   
the role played by gluons in the process.
Complete one-loop results are derived for all structure functions at twist-2.  
With the results presented in this work, more fracture functions, especially 
gluon fracture functions can be extracted from results of relevant experiments.

\section*{Acknowledgements}
The work is supported by National Natural Science Foundation of People’s Republic of China Grants No. 12075299, No. 11821505,  No. 11935017 and by the Strategic Priority Research Program of Chinese Academy of Sciences, Grant No. XDB34000000. 
K. B. Chen is supported by National Natural Science Foundation of China (Nos. 12005122, 11947055), Shandong Province Natural Science Foundation (No. ZR2020QA082), and Youth Innovation Team Program of Higher Education Institutions in Shandong Province (Grant No. 2023KJ126). X. B. Tong  is supported by the Research Council of Finland, the Centre of Excellence in Quark Matter and under the European Union’s Horizon 2020 research and innovation programme by the European Research Council (ERC, grant agreement No. ERC-2018-ADG-835105 YoctoLHC) and by the STRONG-2020 project (grant agreement No. 824093). {The content of this article does not reflect the official opinion of the European Union and responsibility for the information and views expressed therein lies entirely with the authors.}


\begin{thebibliography}{0}
\bibitem{Blumlein:2012bf}
J.~Blumlein,
Prog. Part. Nucl. Phys. \textbf{69}, 28-84 (2013)
doi:10.1016/j.ppnp.2012.09.006
[arXiv:1208.6087 [hep-ph]].

\bibitem{Boussarie:2023izj}
R.~Boussarie, M.~Burkardt, M.~Constantinou, W.~Detmold, M.~Ebert, M.~Engelhardt, S.~Fleming, L.~Gamberg, X.~Ji and Z.~B.~Kang, \textit{et al.}
[arXiv:2304.03302 [hep-ph]].

\bibitem{Boglione:2022gpv}
M.~Boglione \textit{et al.} [Jefferson Lab Angular Momentum (JAM)],
JHEP \textbf{04}, 084 (2022)
doi:10.1007/JHEP04(2022)084
[arXiv:2201.12197 [hep-ph]].

\bibitem{Boglione:2019nwk}
M.~Boglione, A.~Dotson, L.~Gamberg, S.~Gordon, J.~O.~Gonzalez-Hernandez, A.~Prokudin, T.~C.~Rogers and N.~Sato,
JHEP \textbf{10}, 122 (2019)
doi:10.1007/JHEP10(2019)122
[arXiv:1904.12882 [hep-ph]].

\bibitem{Gonzalez-Hernandez:2018ipj}
J.~O.~Gonzalez-Hernandez, T.~C.~Rogers, N.~Sato and B.~Wang,
Phys. Rev. D \textbf{98}, no.11, 114005 (2018)
doi:10.1103/PhysRevD.98.114005
[arXiv:1808.04396 [hep-ph]].

\bibitem{Boglione:2016bph}
M.~Boglione, J.~Collins, L.~Gamberg, J.~O.~Gonzalez-Hernandez, T.~C.~Rogers and N.~Sato,
Phys. Lett. B \textbf{766}, 245-253 (2017)
doi:10.1016/j.physletb.2017.01.021
[arXiv:1611.10329 [hep-ph]].

\bibitem{Mulders:2000jt}
P.~J.~Mulders,
AIP Conf. Proc. \textbf{588}, no.1, 75-88 (2001)
doi:10.1063/1.1413147
[arXiv:hep-ph/0010199 [hep-ph]].

\bibitem{Collins:1989gx}
J.~C.~Collins, D.~E.~Soper and G.~F.~Sterman,
Adv. Ser. Direct. High Energy Phys. \textbf{5}, 1-91 (1989)
doi:10.1142/9789814503266\_0001
[arXiv:hep-ph/0409313 [hep-ph]].

\bibitem{Daleo:2004pn}
A.~Daleo, D.~de Florian and R.~Sassot,
Phys. Rev. D \textbf{71}, 034013 (2005)
doi:10.1103/PhysRevD.71.034013
[arXiv:hep-ph/0411212 [hep-ph]].

\bibitem{Kniehl:2004hf}
B.~A.~Kniehl, G.~Kramer and M.~Maniatis,
Nucl. Phys. B \textbf{711}, 345-366 (2005)
[erratum: Nucl. Phys. B \textbf{720}, 231 (2005)]
doi:10.1016/j.nuclphysb.2005.01.031
[arXiv:hep-ph/0411300 [hep-ph]].

\bibitem{Wang:2019bvb}
B.~Wang, J.~O.~Gonzalez-Hernandez, T.~C.~Rogers and N.~Sato,
Phys. Rev. D \textbf{99}, no.9, 094029 (2019)
doi:10.1103/PhysRevD.99.094029
[arXiv:1903.01529 [hep-ph]].

\bibitem{Collins:1981uk}
J.~C.~Collins and D.~E.~Soper,
Nucl. Phys. B \textbf{193}, 381 (1981)
[erratum: Nucl. Phys. B \textbf{213}, 545 (1983)]
doi:10.1016/0550-3213(81)90339-4

\bibitem{Collins:2004nx}
J.~C.~Collins and A.~Metz,
Phys. Rev. Lett. \textbf{93}, 252001 (2004)
doi:10.1103/PhysRevLett.93.252001
[arXiv:hep-ph/0408249 [hep-ph]].

\bibitem{Ji:2004wu}
X.~d.~Ji, J.~p.~Ma and F.~Yuan,
Phys. Rev. D \textbf{71}, 034005 (2005)
doi:10.1103/PhysRevD.71.034005
[arXiv:hep-ph/0404183 [hep-ph]].

\bibitem{Ji:2004xq}
X.~d.~Ji, J.~P.~Ma and F.~Yuan,
Phys. Lett. B \textbf{597}, 299-308 (2004)
doi:10.1016/j.physletb.2004.07.026
[arXiv:hep-ph/0405085 [hep-ph]].

\bibitem{Collins:2011zzd}
J.~Collins,
Camb. Monogr. Part. Phys. Nucl. Phys. Cosmol. \textbf{32}, 1-624 (2011)
Cambridge University Press, 2023,
ISBN 978-1-00-940184-5, 978-1-00-940183-8, 978-1-00-940182-1
doi:10.1017/9781009401845

\bibitem{Rodini:2023plb}
S.~Rodini and A.~Vladimirov,
[arXiv:2306.09495 [hep-ph]].

\bibitem{Trentadue:1993ka}
L.~Trentadue and G.~Veneziano,
Phys. Lett. B \textbf{323}, 201-211 (1994)
doi:10.1016/0370-2693(94)90292-5

\bibitem{Berera:1995fj}
A.~Berera and D.~E.~Soper,
Phys. Rev. D \textbf{53}, 6162-6179 (1996)
doi:10.1103/PhysRevD.53.6162
[arXiv:hep-ph/9509239 [hep-ph]].

\bibitem{Grazzini:1997ih}
M.~Grazzini, L.~Trentadue and G.~Veneziano,
Nucl. Phys. B \textbf{519}, 394-404 (1998)
doi:10.1016/S0550-3213(97)00840-7
[arXiv:hep-ph/9709452 [hep-ph]].

\bibitem{Collins:1997sr}
J.~C.~Collins,
Phys. Rev. D \textbf{57}, 3051-3056 (1998)
[erratum: Phys. Rev. D \textbf{61}, 019902 (2000)]
doi:10.1103/PhysRevD.61.019902
[arXiv:hep-ph/9709499 [hep-ph]].

\bibitem{Kang:2011mr}
Z.~B.~Kang, B.~W.~Xiao and F.~Yuan,
Phys. Rev. Lett. \textbf{107}, 152002 (2011)
doi:10.1103/PhysRevLett.107.152002
[arXiv:1106.0266 [hep-ph]].

\bibitem{Sun:2013hua}
P.~Sun and F.~Yuan,
Phys. Rev. D \textbf{88}, no.11, 114012 (2013)
doi:10.1103/PhysRevD.88.114012
[arXiv:1308.5003 [hep-ph]].

\bibitem{Benic:2019zvg}
S.~Benic, Y.~Hatta, H.~n.~Li and D.~J.~Yang,
Phys. Rev. D \textbf{100}, no.9, 094027 (2019)
doi:10.1103/PhysRevD.100.094027
[arXiv:1909.10684 [hep-ph]].

\bibitem{Benic:2021gya}
S.~Beni\'c, Y.~Hatta, A.~Kaushik and H.~n.~Li,
Phys. Rev. D \textbf{104}, no.9, 094027 (2021)
doi:10.1103/PhysRevD.104.094027
[arXiv:2109.05440 [hep-ph]].

\bibitem{Benic:2024fvk}
S.~Beni\'c, Y.~Hatta, A.~Kaushik and H.~n.~Li,
[arXiv:2402.02267 [hep-ph]].

\bibitem{Anderle:2016kwa}
D.~Anderle, D.~de Florian and Y.~Rotstein Habarnau,
Phys. Rev. D \textbf{95}, no.3, 034027 (2017)
doi:10.1103/PhysRevD.95.034027
[arXiv:1612.01293 [hep-ph]].

\bibitem{Abele:2021nyo}
M.~Abele, D.~de Florian and W.~Vogelsang,
Phys. Rev. D \textbf{104}, no.9, 094046 (2021)
doi:10.1103/PhysRevD.104.094046
[arXiv:2109.00847 [hep-ph]].

\bibitem{Goyal:2023xfi}
S.~Goyal, S.~O.~Moch, V.~Pathak, N.~Rana and V.~Ravindran,
[arXiv:2312.17711 [hep-ph]].

\bibitem{Bonino:2024qbh}
L.~Bonino, T.~Gehrmann and G.~Stagnitto,
[arXiv:2401.16281 [hep-ph]].

\bibitem{Bergabo:2022zhe}
F.~Bergabo and J.~Jalilian-Marian,
JHEP \textbf{01}, 095 (2023)
doi:10.1007/JHEP01(2023)095
[arXiv:2210.03208 [hep-ph]].

\bibitem{Bergabo:2024ivx}
F.~Bergabo and J.~Jalilian-Marian,
[arXiv:2401.06259 [hep-ph]].

\bibitem{Caucal:2024cdq}
P.~Caucal, E.~Ferrand and F.~Salazar,
[arXiv:2401.01934 [hep-ph]].

\bibitem{Ceccopieri:2008fq}
F.~A.~Ceccopieri and L.~Trentadue,
Phys. Lett. B \textbf{668}, 319-323 (2008)
doi:10.1016/j.physletb.2008.09.001
[arXiv:0805.3467 [hep-ph]].

\bibitem{Ceccopieri:2010zu}
F.~A.~Ceccopieri,
Phys. Lett. B \textbf{703}, 491-497 (2011)
doi:10.1016/j.physletb.2011.08.038
[arXiv:1012.0507 [hep-ph]].

\bibitem{Anselmino:2011ss}
M.~Anselmino, V.~Barone and A.~Kotzinian,
Phys. Lett. B \textbf{699}, 108-118 (2011)
doi:10.1016/j.physletb.2011.03.067
[arXiv:1102.4214 [hep-ph]].

\bibitem{Anselmino:2011bb}
M.~Anselmino, V.~Barone and A.~Kotzinian,
Phys. Lett. B \textbf{706}, 46-52 (2011)
doi:10.1016/j.physletb.2011.10.064
[arXiv:1109.1132 [hep-ph]].

\bibitem{Anselmino:2011vkz}
M.~Anselmino, V.~Barone and A.~Kotzinian,
Phys. Lett. B \textbf{713}, 317-320 (2012)
doi:10.1016/j.physletb.2012.06.003
[arXiv:1112.2604 [hep-ph]].

\bibitem{Chai:2019ykk}
X.~P.~Chai, K.~B.~Chen, J.~P.~Ma and X.~B.~Tong,
JHEP \textbf{10}, 285 (2019)
doi:10.1007/JHEP10(2019)285
[arXiv:1903.00809 [hep-ph]].

\bibitem{Chen:2021vby}
K.~B.~Chen, J.~P.~Ma and X.~B.~Tong,
JHEP \textbf{11}, 038 (2021)
doi:10.1007/JHEP11(2021)038
[arXiv:2108.13582 [hep-ph]].

\bibitem{Guo:2023uis}
Y.~Guo and F.~Yuan,
[arXiv:2312.01008 [hep-ph]].

\bibitem{Hatta:2022lzj}
Y.~Hatta, B.~W.~Xiao and F.~Yuan,
Phys. Rev. D \textbf{106}, no.9, 094015 (2022)
doi:10.1103/PhysRevD.106.094015
[arXiv:2205.08060 [hep-ph]].

\bibitem{Iancu:2021rup}
E.~Iancu, A.~H.~Mueller and D.~N.~Triantafyllopoulos,
Phys. Rev. Lett. \textbf{128}, no.20, 202001 (2022)
doi:10.1103/PhysRevLett.128.202001
[arXiv:2112.06353 [hep-ph]].

\bibitem{Iancu:2022lcw}
E.~Iancu, A.~H.~Mueller, D.~N.~Triantafyllopoulos and S.~Y.~Wei,
JHEP \textbf{10}, 103 (2022)
doi:10.1007/JHEP10(2022)103
[arXiv:2207.06268 [hep-ph]].

\bibitem{Hauksson:2024bvv}
S.~Hauksson, E.~Iancu, A.~H.~Mueller, D.~N.~Triantafyllopoulos and S.~Y.~Wei,
[arXiv:2402.14748 [hep-ph]].

\bibitem{Tong:2023bus}
X.~B.~Tong, B.~W.~Xiao and Y.~Y.~Zhang,
Phys. Rev. D \textbf{109}, no.5, 054004 (2024)
doi:10.1103/PhysRevD.109.054004
[arXiv:2310.20662 [hep-ph]].

\bibitem{Shao:2024nor}
D.~Y.~Shao, Y.~Shi, C.~Zhang, J.~Zhou and Y.~j.~Zhou,
[arXiv:2402.05465 [hep-ph]].

\bibitem{Hatta:2024vzv}
Y.~Hatta and F.~Yuan,
Phys. Lett. B \textbf{854}, 138738 (2024)
doi:10.1016/j.physletb.2024.138738
[arXiv:2403.19609 [hep-ph]].

\bibitem{Chen:2023wsi}
K.~B.~Chen, J.~P.~Ma and X.~B.~Tong,
Phys. Rev. D \textbf{108}, no.9, 9 (2023)
doi:10.1103/PhysRevD.108.094015
[arXiv:2308.11251 [hep-ph]].

\bibitem{Liu:2022wop}
X.~Liu and H.~X.~Zhu,
Phys. Rev. Lett. \textbf{130}, no.9, 9 (2023)
doi:10.1103/PhysRevLett.130.091901
[arXiv:2209.02080 [hep-ph]].

\bibitem{Liu:2023aqb}
H.~Y.~Liu, X.~Liu, J.~C.~Pan, F.~Yuan and H.~X.~Zhu,
Phys. Rev. Lett. \textbf{130}, no.18, 18 (2023)
doi:10.1103/PhysRevLett.130.181901
[arXiv:2301.01788 [hep-ph]].

\bibitem{Li:2023gkh}
X.~L.~Li, X.~Liu, F.~Yuan and H.~X.~Zhu,
Phys. Rev. D \textbf{108}, no.9, L091502 (2023)
doi:10.1103/PhysRevD.108.L091502
[arXiv:2308.10942 [hep-ph]].

\bibitem{Cao:2023oef}
H.~Cao, X.~Liu and H.~X.~Zhu,
Phys. Rev. D \textbf{107}, no.11, 114008 (2023)
doi:10.1103/PhysRevD.107.114008
[arXiv:2303.01530 [hep-ph]].

\bibitem{Cao:2023qat}
H.~Cao, H.~T.~Li and Z.~Mi,
[arXiv:2312.07655 [hep-ph]].

\bibitem{ZEUS:1993vio}
M.~Derrick \textit{et al.} [ZEUS],
Phys. Lett. B \textbf{315}, 481-493 (1993)
doi:10.1016/0370-2693(93)91645-4

\bibitem{Goharipour:2018yov}
M.~Goharipour, H.~Khanpour and V.~Guzey,
Eur. Phys. J. C \textbf{78}, no.4, 309 (2018)
doi:10.1140/epjc/s10052-018-5787-z
[arXiv:1802.01363 [hep-ph]].

\bibitem{Khanpour:2019pzq}
H.~Khanpour,
Phys. Rev. D \textbf{99}, no.5, 054007 (2019)
doi:10.1103/PhysRevD.99.054007
[arXiv:1902.10734 [hep-ph]].

\bibitem{Maktoubian:2019ppi}
A.~Maktoubian, H.~Mehraban, H.~Khanpour and M.~Goharipour,
Phys. Rev. D \textbf{100}, no.5, 054020 (2019)
doi:10.1103/PhysRevD.100.054020
[arXiv:1908.10154 [hep-ph]].

\bibitem{Salajegheh:2022vyv}
M.~Salajegheh, H.~Khanpour, U.~G.~Mei\ss{}ner, H.~Hashamipour and M.~Soleymaninia,
Phys. Rev. D \textbf{106}, no.5, 054012 (2022)
doi:10.1103/PhysRevD.106.054012
[arXiv:2206.13788 [hep-ph]].

\bibitem{Salajegheh:2023jgi}
M.~Salajegheh, H.~Khanpour, U.~G.~Mei\ss{}ner, H.~Hashamipour and M.~Soleymaninia,
Phys. Rev. D \textbf{107}, no.9, 094038 (2023)
doi:10.1103/PhysRevD.107.094038
[arXiv:2301.10284 [hep-ph]].

\bibitem{CLAS:2022sqt}
H.~Avakian \textit{et al.} [CLAS],
Phys. Rev. Lett. \textbf{130}, no.2, 022501 (2023)
doi:10.1103/PhysRevLett.130.022501
[arXiv:2208.05086 [hep-ex]].

\bibitem{Accardi:2023chb}
A.~Accardi, P.~Achenbach, D.~Adhikari, A.~Afanasev, C.~S.~Akondi, N.~Akopov, M.~Albaladejo, H.~Albataineh, M.~Albrecht and B.~Almeida-Zamora, \textit{et al.}
[arXiv:2306.09360 [nucl-ex]].

\bibitem{Boer:2011fh}
D.~Boer, M.~Diehl, R.~Milner, R.~Venugopalan, W.~Vogelsang, D.~Kaplan, H.~Montgomery, S.~Vigdor, A.~Accardi and E.~C.~Aschenauer, \textit{et al.}
[arXiv:1108.1713 [nucl-th]].

\bibitem{Accardi:2012qut}
A.~Accardi, J.~L.~Albacete, M.~Anselmino, N.~Armesto, E.~C.~Aschenauer, A.~Bacchetta, D.~Boer, W.~K.~Brooks, T.~Burton and N.~B.~Chang, \textit{et al.}
Eur. Phys. J. A \textbf{52}, no.9, 268 (2016)
doi:10.1140/epja/i2016-16268-9
[arXiv:1212.1701 [nucl-ex]].

\bibitem{AbdulKhalek:2021gbh}
R.~Abdul Khalek, A.~Accardi, J.~Adam, D.~Adamiak, W.~Akers, M.~Albaladejo, A.~Al-bataineh, M.~G.~Alexeev, F.~Ameli and P.~Antonioli, \textit{et al.}
Nucl. Phys. A \textbf{1026}, 122447 (2022)
doi:10.1016/j.nuclphysa.2022.122447
[arXiv:2103.05419 [physics.ins-det]].

\bibitem{AbdulKhalek:2022hcn}
R.~Abdul Khalek, U.~D'Alesio, M.~Arratia, A.~Bacchetta, M.~Battaglieri, M.~Begel, M.~Boglione, R.~Boughezal, R.~Boussarie and G.~Bozzi, \textit{et al.}
[arXiv:2203.13199 [hep-ph]].

\bibitem{Abir:2023fpo}
R.~Abir, I.~Akushevich, T.~Altinoluk, D.~P.~Anderle, F.~P.~Aslan, A.~Bacchetta, B.~Balantekin, J.~Barata, M.~Battaglieri and C.~A.~Bertulani, \textit{et al.}
[arXiv:2305.14572 [hep-ph]].

\bibitem{Anderle:2021wcy}
D.~P.~Anderle, V.~Bertone, X.~Cao, L.~Chang, N.~Chang, G.~Chen, X.~Chen, Z.~Chen, Z.~Cui and L.~Dai, \textit{et al.}
Front. Phys. (Beijing) \textbf{16}, no.6, 64701 (2021)
doi:10.1007/s11467-021-1062-0
[arXiv:2102.09222 [nucl-ex]].

\bibitem{Bacchetta:2006tn}
A.~Bacchetta, M.~Diehl, K.~Goeke, A.~Metz, P.~J.~Mulders and M.~Schlegel,
JHEP \textbf{02}, 093 (2007)
doi:10.1088/1126-6708/2007/02/093
[arXiv:hep-ph/0611265 [hep-ph]].

\bibitem{Graudenz:1994dq}
D.~Graudenz,
Nucl. Phys. B \textbf{432}, 351-376 (1994)
doi:10.1016/0550-3213(94)90606-8
[arXiv:hep-ph/9406274 [hep-ph]].

\bibitem{deFlorian:1995fd}
D.~de Florian, C.~A.~Garcia Canal and R.~Sassot,
Nucl. Phys. B \textbf{470}, 195-210 (1996)
doi:10.1016/0550-3213(96)00159-9
[arXiv:hep-ph/9510262 [hep-ph]].

\bibitem{Daleo:2003jf}
A.~Daleo and R.~Sassot,
Nucl. Phys. B \textbf{673}, 357-384 (2003)
doi:10.1016/j.nuclphysb.2003.09.007
[arXiv:hep-ph/0309073 [hep-ph]].

\bibitem{Daleo:2003xg}
A.~Daleo, C.~A.~Garcia Canal and R.~Sassot,
Nucl. Phys. B \textbf{662}, 334-358 (2003)
doi:10.1016/S0550-3213(03)00334-1
[arXiv:hep-ph/0303199 [hep-ph]].

\bibitem{Diehl:2005pc}
M.~Diehl and S.~Sapeta,
Eur. Phys. J. C \textbf{41}, 515-533 (2005)
doi:10.1140/epjc/s2005-02242-9
[arXiv:hep-ph/0503023 [hep-ph]].

\bibitem{tHooft:1972tcz}
G.~'t Hooft and M.~J.~G.~Veltman,
Nucl. Phys. B \textbf{44}, 189-213 (1972)
doi:10.1016/0550-3213(72)90279-9

\bibitem{Breitenlohner:1977hr}
P.~Breitenlohner and D.~Maison,
Commun. Math. Phys. \textbf{52}, 11-38 (1977)
doi:10.1007/BF01609069

\bibitem{Ellis:1996mzs}
R.~K.~Ellis, W.~J.~Stirling and B.~R.~Webber,
Camb. Monogr. Part. Phys. Nucl. Phys. Cosmol. \textbf{8}, 1-435 (1996)
Cambridge University Press, 2011,
ISBN 978-0-511-82328-2, 978-0-521-54589-1
doi:10.1017/CBO9780511628788

\bibitem{Gluck:1995yr}
M.~Gluck, E.~Reya, M.~Stratmann and W.~Vogelsang,
Phys. Rev. D \textbf{53}, 4775-4786 (1996)
doi:10.1103/PhysRevD.53.4775
[arXiv:hep-ph/9508347 [hep-ph]].

\bibitem{Vogelsang:1995vh}
W.~Vogelsang,
Phys. Rev. D \textbf{54}, 2023-2029 (1996)
doi:10.1103/PhysRevD.54.2023
[arXiv:hep-ph/9512218 [hep-ph]].

\bibitem{deFlorian:1997ie}
D.~de Florian, O.~A.~Sampayo and R.~Sassot,
Phys. Rev. D \textbf{57}, 5803-5810 (1998)
doi:10.1103/PhysRevD.57.5803
[arXiv:hep-ph/9711440 [hep-ph]].

\bibitem{Altarelli:1977zs}
G.~Altarelli and G.~Parisi,
Nucl. Phys. B \textbf{126}, 298-318 (1977)
doi:10.1016/0550-3213(77)90384-4
\end{thebibliography}
\end{document}